\def\year{2022}\relax
\documentclass[letterpaper]{article} 
\usepackage{aaai22}  
\usepackage{times}  
\usepackage{helvet}  
\usepackage{courier}  
\usepackage[hyphens]{url}  
\usepackage{graphicx} 
\usepackage{natbib}  
\usepackage{caption} 
\DeclareCaptionStyle{ruled}{labelfont=normalfont,labelsep=colon,strut=off} 
\usepackage{CJK}
\usepackage{booktabs}
\usepackage{threeparttable}
\usepackage{amsmath}
\usepackage{makecell}
\usepackage{multicol}
\usepackage{amsthm,amsmath,amssymb}
\usepackage{mathrsfs}
\usepackage{subfig}
\usepackage{float}
\usepackage{bm}
\usepackage{color}
\usepackage{multirow}
\usepackage{algorithm}
\usepackage{algorithmic}

\usepackage{newfloat}
\usepackage{listings}
\floatstyle{ruled}
\newfloat{listing}{tb}{lst}{}
\floatname{listing}{Listing}
\title{Mind the Gap:  Cross-Lingual Information Retrieval \\ with Hierarchical Knowledge Enhancement}
\author{
    Fuwei Zhang\textsuperscript{\rm 1, \rm 2}, Zhao Zhang\textsuperscript{\rm 3,}\thanks{Corresponding author: Zhao Zhang}, Xiang Ao\textsuperscript{\rm 1, \rm2, \rm 4}, Dehong Gao\textsuperscript{\rm 5}, Fuzhen Zhuang\textsuperscript{\rm 6, \rm 7}, \\Yi Wei\textsuperscript{\rm 5}, Qing He\textsuperscript{\rm 1, \rm 2}
}
\affiliations{
    \textsuperscript{\rm 1} Key Lab of Intelligent Information Processing of Chinese Academy of Sciences (CAS),\\ Institute of Computing Technology, CAS, Beijing 100190, China \\
    \textsuperscript{\rm 2} University of Chinese Academy of Sciences, Beijing 100049, China \\
    \textsuperscript{\rm 3} Institute of Computing Technology, Chinese Academy of Sciences, Beijing 100190, China  \\
    \textsuperscript{\rm 4} Institute of Intelligent Computing Technology, Suzhou, CAS  \ \ \ \ \ \  
    \textsuperscript{\rm 5} Alibaba Group, Hangzhou, China \\
    \textsuperscript{\rm 6} Institute of Artificial Intelligence, Beihang University, Beijing 100191, China \\
    \textsuperscript{\rm 7} SKLSDE, School of Computer Science, Beihang University, Beijing 100191, China

    \{zhangfuwei20g, zhangzhao2021, aoxiang, heqing\}@ict.ac.cn, zhuangfuzhen@buaa.edu.cn, \\ \{dehong.gdh, yi.weiy\}@alibaba-inc.com
}

\usepackage{bibentry}

\begin{document}

\maketitle

\begin{abstract}

\begin{quote}

Cross-Lingual Information Retrieval~(CLIR) aims to rank the documents written in a language different from the user's query. 
The intrinsic gap between different languages is an essential challenge for CLIR. 
In this paper, we introduce the multilingual knowledge graph~(KG) to the CLIR task due to the sufficient information of entities in multiple languages. It is regarded as a ``silver bullet'' to simultaneously perform explicit alignment between queries and documents and also broaden the representations of queries.
And we propose a model named \textit{CLIR with \underline{hi}erarchical \underline{k}nowledge \underline{e}nhancement} (HIKE) for our task. The proposed model encodes the textual information in queries, documents and the KG with multilingual BERT, and incorporates the KG information in the query-document matching process with a hierarchical information fusion mechanism. Particularly, HIKE first integrates the entities and their neighborhood in KG into query representations with a  knowledge-level fusion, then combines the knowledge from both source and target languages to further mitigate the linguistic gap with a language-level fusion. Finally, experimental results demonstrate that HIKE achieves substantial improvements over state-of-the-art competitors.


\end{quote}
\end{abstract}

\section{Introduction}

The escalation of globalization burgeons the great demand for Cross-Lingual Information Retrieval~(CLIR), which has broad applications such as cross-border e-commerce, cross-lingual question answering, and so on \cite{eco,ruckle2019improved, xu2021artificial}. 
Informally, given a query in one language, CLIR is a document retrieval task that aims to rank the candidate documents in another language according to the relevance between the search query and the documents. 

Most existing solutions to tackle the CLIR task are built upon machine translation~\cite{dwivedi2016survey} systems~(also known as MT systems). One technical route is to translate either the query or the document to the same language as the other side~\cite{dic-trans1,dic-trans2,cor-trans1,doc-trans1}. The other is to translate both the query and the document to the same intermediate language~\cite{kishida2003two}, e.g. English. After aligning the language of the query and documents, monolingual retrieval is performed to accomplish the task. Hence, the performance of the MT systems and the error accumulations may render them inefficient in CLIR. 

Recent studies strive to model CLIR with deep neural networks that encode both query and document into a shared space rather than using MT systems~\cite{zhang2019improving,share-repre,hui2018co,eco}. Though these approaches achieve some remarkable successes, the intrinsic differences between different languages still exist due to the implicit alignment of these methods.
Meanwhile, the query is not very long, leading the lack of information while matching with candidate documents. 

\begin{figure}[t]
	  \centering
	  \setlength{\abovecaptionskip}{2pt}
	  \setlength{\belowcaptionskip}{-5pt}
      \includegraphics[width=\linewidth]{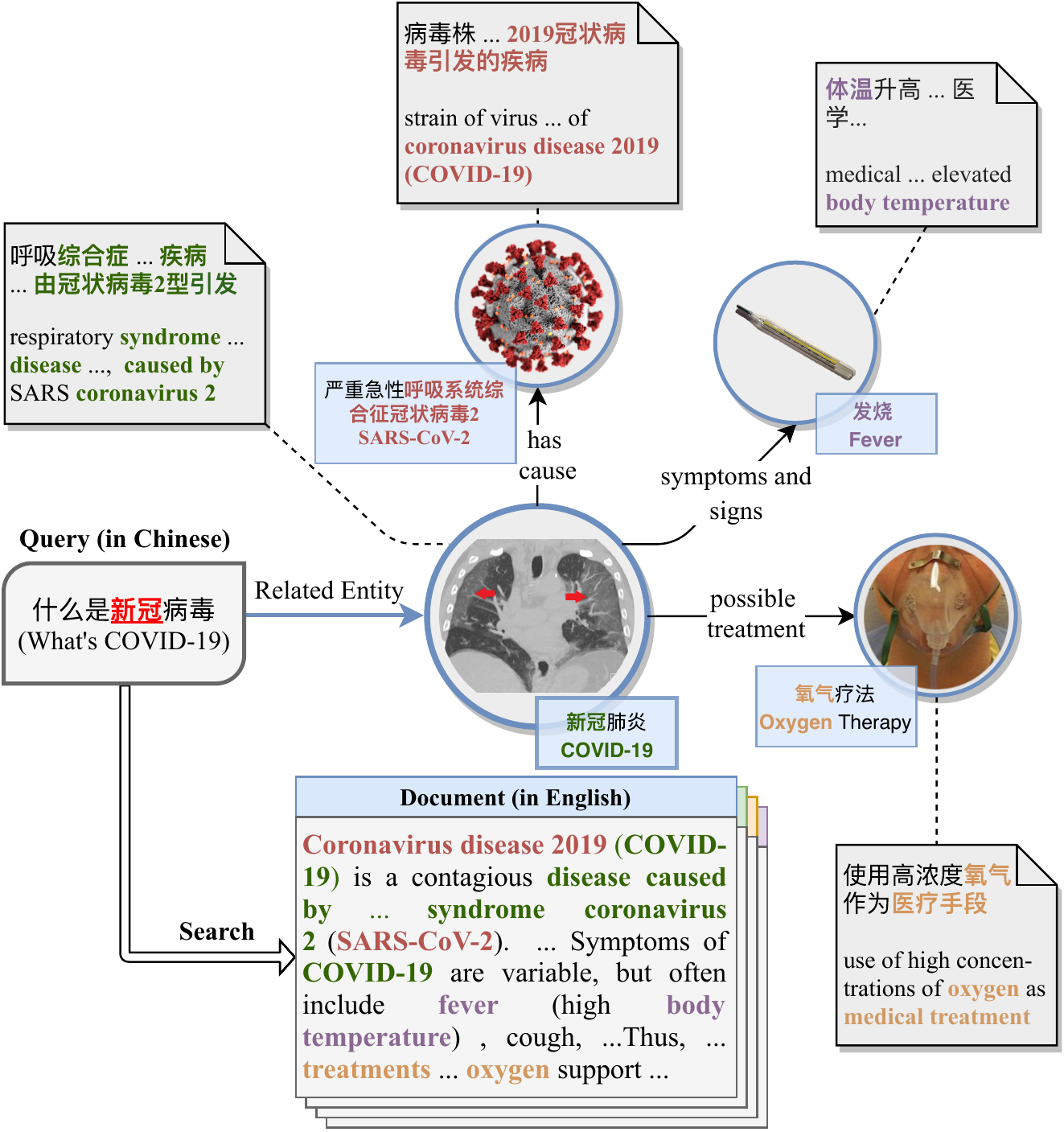}
      \caption{A toy example for utilizing the multilingual KG for CLIR. The query is in Chinese and the documents are in English. In the query, we give an English translation for better understanding. The entities are denoted in circles. The dotted black line presents the descriptions of an entity. The solid black arrow presents relations between entities. The solid blue arrow shows the related entity of the given query. The hollow arrow presents the documents of the query. The entities and corresponding descriptions in KG are bilingual.} 
      \label{fig_intr}
\end{figure}

To tackle these issues, 
we aim to find a ``silver bullet'' to simultaneously perform \emph{explicit alignment} between queries and documents and \emph{broaden} the information of queries. 
The multilingual knowledge graph~(KG), e.g. Wikidata~\cite{vrandevcic2014wikidata}, is our answer. 
As a representative multilingual KG,  Wikidata\footnote{\url{https://www.wikidata.org/wiki/Wikidata:Main_Page}} includes more than 94 million entities and 2 thousand kinds of relations, and most of the entities in Wikidata have multilingual aligned names and descriptions\footnote{ More than 260 languages are supported now.}. With such an external source of knowledge, we can build an explicit bridge between the source language and target language on the premise of the given query information. 
For example, Figure~\ref{fig_intr} exhibits a query ``\begin{CJK*}{UTF8}{gbsn}新冠病毒\end{CJK*}'' in Chinese (``COVID-19'' in English) and candidate documents in English. 
Through the multilingual KG, we could link ``\begin{CJK*}{UTF8}{gbsn}新冠病毒\end{CJK*}'' to its aligned entity in English, i.e. ``COVID-19'', and then extend to some related neighbors, such as ``Fever'', ``SARS-CoV-2'' and ``Oxygen Therapy''. 
Both the aligned entity and the local neighborhood might contribute to extend the insufficient query and fill in the linguistic gap between the query and documents.

Along this line, we adopt the multilingual KG as an external source to facilitate CLIR and propose a \textbf{HI}erarchical \textbf{K}nowledge \textbf{E}nhancement~(HIKE for short) mechanism to fully integrate the relevant knowledge. 
Indeed, queries are usually short but rich in entities. HIKE establishes a link between queries and multilingual KG through the entities mentioned in queries, and makes full use of the semantic information of entities and their neighborhood in KG with a hierarchical information fusion mechanism.  Specifically, a knowledge-level fusion integrates the information in  each individual language in the KG, and a language-level fusion combines the integrated information from different languages. The multilingual KG provides valuable information, which helps to reduce the disparity  between different languages and is beneficial to the matching process over queries and documents.

To summarize, the contributions are as follows.
\begin{itemize}
    \item We adopt the external multilingual KG not only as an enhancement for sparse queries but also as an explicit bridge mitigating the gap between the query and the document in CLIR. To the best of our knowledge, this is the first work that utilizes multilingual KG for the neural CLIR task. 
    
    \item We propose HIKE that makes full use of the entities mentioned in queries as well as the local neighborhoods in the multilingual KG for improving the performance in CLIR. HIKE contains a hierarchical information fusion mechanism to resolve the sparsity in queries and perform easier matching over the query-document pairs. 
   
    \item Extensive experiments on a number of benchmark datasets in four languages~(English, Spanish, French, Chinese) validate the effectiveness of HIKE against state-of-the-art baselines. 
\end{itemize}


\section{Related Work}
%
%


Current information retrieval models for cross-lingual tasks can be categorized into two groups: (i) translation-based approaches~\cite{nie2010cross, zbib2019neural} and (ii) semantic alignment approaches~\cite{bai2010learning,sokolov2013boosting}. 

Early works mainly focus on translation-based models. One way is to translate queries to the target language of documents~\cite{query-trans-1}, or to translate the documents or corpus to the same language as queries~\cite{doc-trans-1,doc-trans-2}. The other is to translate both queries and documents to the same intermediate language, e.g. English~\cite{kishida2003two}.  
In both cases, they aim to simplify the process and use the monolingual information retrieval methods to solve the CLIR problem. 

Recently, with the development of deep neural networks, semantic alignment approaches, which directly tackle the CLIR tasks without the translation process, have gained much attention. These methods align queries and documents into the same space with probabilistic or neural network methods and perform query-document matching in the aligned space.
\citet{sokolov2013boosting} proposed a method about learning bilingual n-gram correspondences from relevance rankings. \citet{share-repre} presented a simple yet effective method using shared representations across CLIR models trained in different language pairs.
The release of BERT~\cite{bert} leads to breakthroughs in various NLP tasks~\cite{jiang2020cross}, including document ranking tasks.
Thus Contextualized Embeddings for Document Ranking~(CEDR)~\cite{cedr} is an effective method for using BERT to enhance the current prevalent neural ranking models, such as KNRM~\cite{knrm}, PACRR~\cite{pacrr} and DRMM~\cite{drmm}. \citet{clirmatrix} utilized a multilingual version of BERT~(a.k.a multilingual BERT or mBERT) to conduct the CLIR task. These BERT-based neural ranking models achieve the state-of-the-art results compared with other models.

 Besides, due to the fast-growing scale of KGs such as Wikidata~\cite{vrandevcic2014wikidata} and DBpedia~\cite{auer2007dbpedia}, some researches focus on using high-quality KGs as extra knowledge to perform the information retrieval task. \citet{word-entity} presented a word-entity duet framework for utilizing KGs in ad-hoc retrieval. Entity-Duet Neural Ranking Model~(EDRM)~\cite{entityduet}, which introduces KGs to neural search systems, represents queries and documents by their word and entity annotations. 
Despite the popularity of KG for information retrieval, the works on the topic of KG for CLIR are rarely found. \citet{zhang2016xknowsearch} introduced KG to CLIR systems using the standard similarity measures for document ranking. However, this work does not use neural network models. To the best of our knowledge, our work is the first work that incorporates multilingual KG information for the neural CLIR task.

\section{Methodology}
In this section, we illustrate the overall framework of our HIKE model, including the model architecture and the detailed description of model components. 

\begin{figure*}[t]
	  \centering
	 
	  \setlength{\abovecaptionskip}{2pt}
	  \setlength{\belowcaptionskip}{-10.0pt}
      \includegraphics[width=\linewidth]{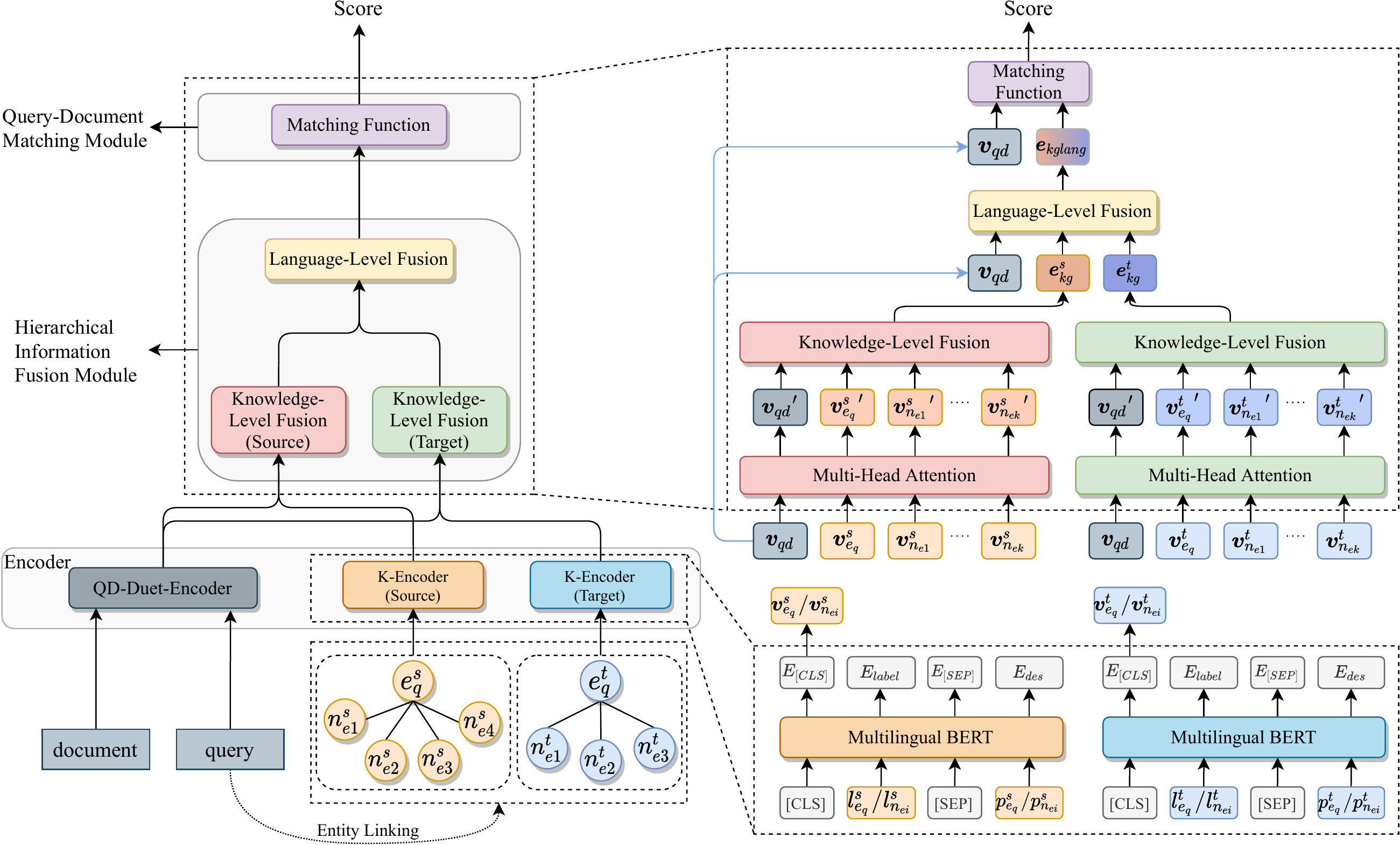}
      \caption{The overall framework of HIKE. The left part is the general architecture, and the right part is the detailed illustration.
    } 
      \label{model}
\end{figure*}

\subsection{Notations}
CLIR is a retrieval task in which search queries and candidate documents are written in different languages. 
Since search queries are usually short but rich in entities, HIKE establishes a connection between CLIR and the multilingual KG via the entities mentioned in queries, and leverage the KG information through these entities and their local neighborhood in KG. Specifically, for each entity, we obtain the following information from the multilingual KG: (i) entity label\footnote{In some large-scale KGs like Wikidata~\cite{vrandevcic2014wikidata} and DBpedia~\cite{auer2007dbpedia}, the name of an entity is denoted as its label.}, (ii) entity description, (iii) labels of neighboring entities, and (iv) descriptions of neighboring entities. It is worth noting that all the information in the KG is multilingual, and the information in different languages is aligned.
We leverage the above information to facilitate the CLIR task.
Given a query $q$ and a document $d$. We present an entity $e_q \in \mathcal{E}$ and the $i$-th neighboring entity $n_{ei} \in \mathcal{E}$, where $\mathcal{E}$ is the entity set in KG. Both the entity and neighboring entities have two information for incorporating: labels and descriptions. Furthermore, for a specific bilingual information retrieval task, the label and description of $e_q$ can be described as $l_{e_q}^r$ and $p_{e_q}^r$, respectively. The label and the description of $n_{ei}$ can be descried as $l_{n_{ei}}^r$ and $p_{n_{ei}}^r$, where $r \in \{s, t\}$ indicates the source language or target language. All these information, including $q$, $d$, $l_{e_q}^r$, $p_{e_q}^r$, $l_{n_{ei}}^r$ and  $p_{n_{ei}}^r$, is composed of a sequence of tokens.


\subsection{Model Architecture}

HIKE incorporates the multilingual semantic information of the entities and their local neighborhoods from KG into the current CLIR model. 
The overall architecture of HIKE is shown in Figure 2. 
HIKE consists of three modules: an encoder module, a hierarchical information fusion module and a query-document matching module. 
Specifically, in the encoder module, HIKE utilizes multilingual BERT to embed the queries, documents, and semantic information from KG into low-dimensional vectors. 
Thus the encoder outputs the embeddings to the hierarchical information fusion module, and the latter combines the information from KG into queries and expedites the matching with documents.
Particularly, the knowledge-level (first-level) fusion integrates the information in KG, using the multi-head attention mechanism~\cite{attention}. We use two individual knowledge-level fusion modules to extract features from source and target languages. And then, the language-level~(second-level) fusion integrates two representations of an entity in source and target languages through a multi-layer perceptron. After the hierarchical information fusion mechanisms, we utilize a matching model to get the relevance score of the query-document pair. 
The higher the score, the more relevant the query and the document are.

\subsection{Encoder}

The encoder aims to embed the tokens from queries, documents, entities and neighboring entities.  It consists of two parts: Query and Document Duet Encoder~(QD-Duet-Encoder) and Knowledge  Encoder~(K-Encoder). QD-Duet-Encoder embeds a query-document pair to a $d$-dimensional vector. And K-Encoder transforms the label and description of an entity into another $d$-dimension vector.


\noindent \textbf{QD-Duet-Encoder} concatenates the tokens from queries and documents into one sequence, using  [CLS] and [SEP] as meta-tokens. [CLS] is a special symbol added in front of every input example, and [SEP] is a special separator token~\cite{bert}. And then the encoder sums the token embedding, segment embedding, positional embedding for each token to get the input embedding and computes the output embedding that represents the semantic and matching information of a query-document pair. Embedding query and document together can make the ranking model benefit from deep semantic information from BERT in addition to individual contextualized token matching~\cite{cedr}. For a given query $q$ and document $d$, we have an output from QD-Duet-Encoder as shown in Equation (\ref{qd-encode}). $\bm{v}_{qd}$ is the [CLS] embedding of the output.
\begin{equation}
\begin{aligned}
    \bm{v}_{qd} =  {\mbox{QD-Duet-Encoder}}( \{\mbox{[CLS]}, q, \mbox{[SEP]}, d\}),
    \label{qd-encode}
\end{aligned}
\end{equation}
where $\mbox{QD-Duet-Encoder}(\cdot)$ is a multilingual BERT model\footnote{We used BERT-base, multilingual cased.} and  $\{\cdot,\cdot\}$ means concatenating two sequences of tokens to one sequence.

\noindent  \textbf{K-Encoder} aims to embed the knowledge information from entities or  neighboring entities in two languages to a feature vector. Inspired by the advantages of embedding the query and document together, we use [CLS] and [SEP] to concatenate the label and the description of an entity to obtain the embedding. Suppose there are $k$ neighboring entities, we denote the set of neighboring entity labels as $\mathcal{N}_l^r = \{l_{n_{e1}}^r, l_{n_{e2}}^r, \cdots, l_{n_{ek}}^r\}$ and the descriptions as $\mathcal{N}_p^r = \{p_{n_{e1}}^r, p_{n_{e2}}^r, \cdots, p_{n_{ek}}^r\}$. All these entities are fed into K-Encoder to compute a feature embedding of the entity as
\begin{equation}
\begin{aligned}
  \bm{v}_{e_q}^r & = \mbox{K-Encoder}(\{\mbox{[CLS]}, l_{e_q}^r, \mbox{[SEP]}, p_{e_q}^r\}), \\
   \bm{v}_{n_{ei}}^r & = \mbox{K-Encoder}(\{\mbox{[CLS]}, l_{n_{ei}}^r, \mbox{[SEP]}, p_{n_{ei}}^r\}), 
   \label{k-encode}
\end{aligned}
\end{equation}
where $i=1,2,\cdots,k$. $\mbox{K-Encoder}(\cdot)$ is also a multilingual BERT. 
$r \in \{s, t\}$ denotes that the parameter is for source and target languages, respectively.
We sort the neighboring entities in descending order according to their relevance to the central entity and select  top $k$ neighboring entities to obtain $\bm{v}_{n_{ei}}^r$, where $k$ is a hyper-parameter. Specifically, we first run the popular KG embedding model TransE~\cite{TransE} to get the embeddings of entities, and then calculate the cosine similarity between two entities as the relevance score.
$\bm{v}_{e_q}^r$ and $\bm{v}_{n_{ei}}^r$ are the [CLS] embedding of the entity and the $i$-th neighboring entity, respectively.  The set of feature vectors of neighboring entities is  $\mathcal{N}^r = \{\bm{v}_{n_{e1}}^r, \bm{v}_{n_{e2}}^r, \cdots,\bm{v}_{n_{ek}}^r\}$. $\bm{v}_{qd}$, $\bm{v}_{e_q}^r$ and $\mathcal{N}^r$ will be treated as the inputs of the fusion module 
in the next subsection.

\subsection{Hierarchical Information Fusion}

In this section, we detail the hierarchical information fusion module, which is a two-level fusion mechanism, comprising knowledge-level fusion and language-level fusion.

\noindent \textbf{Knowledge-Level Fusion} contains two modules: a multi-head self-attention mechanism and an information aggregator. With the help of both two modules, our model can learn a wealth of similar semantic information among the entity, neighboring entities and query-doc pair. In the self-attention mechanism,  $\bm{v}_{qd}$, $\bm{v}_{e_q}^r$ and $\mathcal{N}^r$ are gathered together and fed into the attention module to calculate the attention values. The input matrix $\bm{E}^r$ is denoted as:
\begin{equation}
    \begin{aligned}
        \bm{E}^r = (\bm{v}_{qd} \odot \bm{v}_{e_q}^r \odot \bm{v}_{n_{e1}}^r \odot \bm{v}_{n_{e2}}^r \odot \cdots \odot \bm{v}_{n_{ek}}^r),
        \label{input}
    \end{aligned}
\end{equation}
where $\odot$ is an operation that stacks row vectors into a matrix. 

$\bm{E}^r$ contains the embeddings from query, document, entity and the local neighborhood of the entity. To encapsulate more valuable information, we utilize the multi-head attention mechanism~\cite{attention} to learn better latent semantic information. The self-attention module takes three inputs~(the query, the key, and the value), which are denoted as $\bm{Q}$, $\bm{K}$, $\bm{V} \in \mathbb{R}^{(2+k) \times d}$ ($d$ is the embedding size) respectively. 
To be specific, we only discuss the $j$-th head of the multi-head attention mechanism. First, the self-attention model uses each embedding in $\bm{E}^r$ to get the query $\bm{Q}^j$, key $\bm{K}^j$ and value $\bm{V}^j$ through a linear transformation layer. Then the model goes on using each embedding in the query to attend each embedding in the key through the scaled dot-product attention mechanism \cite{attention}, and gets the attention score. Finally, the obtained attention score is applied upon the value $\bm{V}^j$ to calculate a new representation of $\mbox{Att}(\bm{Q}^j, \bm{K}^j, \bm{V}^j)$, which is formulated as: 
\begin{equation}
    \mbox{Att}(\bm{Q}^j, \bm{K}^j, \bm{V}^j) = \mathrm{softmax}(\frac{\bm{Q}^j \cdot (\bm{K}^j)^T}{\sqrt{d}})\cdot \bm{V}^j.
\end{equation}
Therefore, each row of $\mbox{Att}(\bm{Q}^j, \bm{K}^j, \bm{V}^j)$ is capable of incorporating the semantic information from the rows in $\bm{V}^j$. Furthermore, a layer normalization operation~\cite{ba2016layer} is applied to the output of attention model to obtain the representation of the $j$-th head $\bm{H}^j = \mathrm{LayerNorm(}\mbox{Att}(\bm{Q}^j, \bm{K}^j, \bm{V}^j))$. Next, we pack the multi-head information using the following operation: 
\begin{equation}
     \mbox{Multi-Head}(\bm{Q}, \bm{K}, \bm{V}) = (\bm{H}^1||\bm{H}^2||\cdots|| \bm{H}^m)\bm{W}_H,
\end{equation}
where $\bm{W}_H \in \mathbb{R}^{md \times d}$ is a parameter matrix and $m$ is the number of heads. 

Accordingly, we obtain the representation after the multi-head attention $\bm{M}^r = ({\bm{v}_{qd}}^{\prime} \odot {\bm{v}_{e_q}^r}^{\prime} \odot {\bm{v}_{n_{e1}}^r}^{\prime} \odot {\bm{v}_{n_{e2}}^r}^{\prime} \odot \cdots \odot {\bm{v}_{n_{ek}}^r}^{\prime}) = \mbox{Multi-Head}(\bm{Q}, \bm{K}, \bm{V})  \in \mathbb{R}^{(2+k)\times d}$, where $r \in \{s, t\}$ denotes that the parameter is for source and target languages respectively.
${\bm{v}_{qd}}^{\prime}$, ${\bm{v}_{e_q}^r}^{\prime}$ and ${\bm{v}_{n_{ei}}^r}^{\prime} (i = 1,2,\dots,k)$ represent the output vectors of multi-head self attention. Finally, we use an information aggregator which consists of a linear transformation layer as Equation (\ref{aggre}) to compute the final representation of the knowledge-level features. 
\begin{equation}
     \bm{e}_{kg}^r = \mathrm{Tanh}(\bm{W}_K \cdot \mathrm{vec}(\bm{M}^r) +\bm{b}_K),
     \label{aggre}
\end{equation}
where $\mathrm{vec}(\cdot)$ is a vectorization function that concatenates each row of a matrix as a long vector. $\bm{W}_K \in \mathbb{R}^{d \times (2+k)d}$ is a parameter matrix and $\bm{b}_K$ is a $d$-dimension vector.
$\bm{e}_{kg}^r$ incorporates the deep semantic information from the KG.

\noindent \textbf{Language-Level Fusion} combines the query-document pair information with $\bm{e}_{kg}^s$ and $\bm{e}_{kg}^t$, which are obtained from the knowledge-level fusion. We use the $\bm{v}_{qd}$ as guidance in the fusion processing, which is donated in blue arrow in Figure~\ref{model}. And then, these embeddings are combined by a linear transformation layer which uses $\mathrm{Tanh}$ as the activation function to generate a unified representation as: 
\begin{equation}
    \bm{e}_{kglang} =\mathrm{Tanh}[\bm{W}_L(\bm{v}_{qd}||\bm{e}^{s}_{kg}||\bm{e}^{t}_{kg}) + \bm{b}_L],
\end{equation}
where $s$ and $t$ represent the source and target languages.
$\bm{W}_L \in \mathbb{R}^{d \times 3d}$ and $\bm{b}_L \in \mathbb{R}^d$ are parameters.
$\bm{e}_{kglang}$ is the unified embedding that incorporates the information from queries, documents, and the multilingual KG.

\subsection{Matching Function}

Finally, HIKE uses the matching function to obtain the score of a query-document pair.
Particularly, $\bm{v}_{qd}$ and $\bm{e}_{kglang}$ will be concatenated and fed into another linear layer to obtain the relevant ranking score of the query-document pair: 
\begin{equation}
    f(q, d) = \mathrm{Softmax}[\bm{W}_S(\bm{v}_{qd}||\bm{e}_{kglang}) + b_S],
\end{equation}
where $f(q, d)$ is the ranking score between the query and document.
$\bm{W}_S \in \mathbb{R}^{1 \times 2d}$ and $b_S \in \mathbb{R}^1$ are parameters.
And $\mathrm{Softmax}$ is an activate function to convert the results into the probability over different classes.

In the training stage, we use standard pairwise hinge loss to train the model as shown in Equation (\ref{loss}).
\begin{equation}
    \mathcal{L} = \sum\limits_{d \in D_q^+} \sum\limits_{d' \in D_q^-}[1 - f(q, d) + f(q, d')]_+.
    \label{loss}
\end{equation}
$D_q^+$ and $D_q^-$ are the set of relevant documents and irrelevant documents of the query $q$ , and $[\cdot]_+ = \text{max}(0, \cdot)$.

\section{Experiment Methodology}
In this section, we describe the details of our experiments, including the dataset, the multilingual KG, baselines, evaluation metrics and implementation details.

\subsection{Dataset} 
We evaluate the HIKE model in a public CLIR dataset CLIRMatrix \cite{clirmatrix}.
Specifically, we use the MULTI-8 set in CLIRMatrix, in which queries and documents are jointly aligned in 8 different languages. The dataset is mined from 49 million unique queries and 34 billion (query, document, relevance label) triplets. The relevance label $\in \{0, 1, 2, 3, 4, 5, 6\}$ indicates the relevance of the query-document pair. The higher the value, the more relevant the query-document pair is.
In MULTI-8, queries remain the same no matter what the language of documents is. For instance, three language pairs English-Spanish, English-French and English-Chinese in MULTI-8 share the same queries. Furthermore, we choose four widely used languages in the world to conduct the bilingual information retrieval tasks, including English~(EN), French~(FR), Spanish~(ES) and Chinese~(ZH). Thus there are 12 language pairs in the dataset for training, validation and testing. The training sets of every language pair contain 10,000 queries, while the validation and the test sets contain 1,000 queries. Meanwhile,  the number of candidate documents for each query is 100. We use the test1 set in MULTI-8 as our test set to verify the model performance. The statistics of the datasets are summarized in Table \ref{data}.
\begin{table}[h]
    \centering
    \setlength{\abovecaptionskip}{0.1cm}
    \setlength{\belowcaptionskip}{-0.5cm}
    \begin{tabular}{cccc}
    \toprule
        Dataset &  train & valid & test \cr
        \midrule
        $\{s \rightarrow t\}$ & 10000 & 1000 & 1000 \cr
        \bottomrule
    \end{tabular}
    \caption{Statistic of datasets. Here $ s, t \in \{\mbox{EN}, \mbox{ES},\mbox{FR},$ $ \mbox{ZH}\}$ and $s \neq t$.}
    \label{data}
\end{table}
\begin{table}[h]
    \centering
    \setlength{\abovecaptionskip}{0.1cm}
    \setlength{\belowcaptionskip}{-0.5cm}
    \fontsize{8}{6}\selectfont    
    \setlength\tabcolsep{3.0pt}
    \begin{tabular}{lcccccc}
    \toprule
        &EN-ES &  EN-FR & EN-ZH & ES-EN & ES-FR & ES-ZH \cr
        \midrule
        source language & \multicolumn{3}{c}{7.11}&\multicolumn{3}{c}{6.53}\cr
        \cmidrule(lr){2-4} 
        \cmidrule(lr){5-7}
        target language & 6.15&6.34&4.86&7.37&6.73&5.21 \cr
        \bottomrule
        \toprule
        & FR-EN &  FR-ES & FR-ZH & ZH-EN & ZH-ES & ZH-FR \cr
        \midrule
        source language & \multicolumn{3}{c}{6.41}&\multicolumn{3}{c}{4.95} \cr
         \cmidrule(lr){2-4} 
        \cmidrule(lr){5-7}
        target language & 7.11&6.19&4.93&7.02&6.13&6.33 \cr
        \bottomrule
    \end{tabular}
    \caption{
    Average number of golden neighboring entities. ``Golden'' means the neighboring entities have both the description and the label in a specific language of the queries. The source language is on the left of the connector ``-'', while the target language is on the right. 
    }
    \label{know}
\end{table}
\begin{table*}[t] 
		\centering 
		\setlength{\abovecaptionskip}{0.1cm}
		\setlength{\belowcaptionskip}{-8pt}
		\fontsize{8}{7}\selectfont    

		\begin{threeparttable} 
			\begin{tabular}{p{1.8cm}<{\centering}p{1.5cm}p{1.8cm}<{\centering}p{1.8cm}<{\centering}p{1.8cm}<{\centering}p{1.8cm}<{\centering}p{1.8cm}<{\centering}p{1.8cm}<{\centering}}
				\toprule         
				
				\multirow{2}[2]{*}{\bf{Language Pair}}&  \multirow{2}[2]{*}{\bf Metrics}  & \multicolumn{6}{c}{\bf Models} \cr

				\cmidrule(lr){3-8}
				& & Vanilla BERT & CEDR-DRMM & CEDR-KNRM & CEDR-PACRR & HIKE$^{-}$  & HIKE \cr
				\midrule 
				\multirow{3}{*}{EN-ES} & NDCG@1 & 75.82 & 73.55&75.40&77.28&80.05 &\bf 83.81$^*$\cr
				  & NDCG@5 & 80.08&79.19&80.30&80.69& 82.63 &\textbf{84.05}$^*$\cr
				  & NDCG@10 & 83.36&82.55&83.47&83.42& 85.14& \textbf{86.18}$^*$\cr

				 \cmidrule(lr){2-8}
				\multirow{3}{*}{EN-FR}& NDCG@1 & 76.92&74.63&71.40&78.33&80.05 &\bf 82.93$^*$\cr
				  & NDCG@5 & 78.99&78.27&78.53&80.90&81.21 &\textbf{83.43}$^*$\cr
				  & NDCG@10 & 82.02&81.01&81.89&83.40&83.20 &\bf{85.22}$^*$\cr

				\cmidrule(lr){2-8}
				\multirow{3}{*}{EN-ZH}& NDCG@1 & 68.98&70.33&76.60&75.10&72.25 &\bf 78.16$^*$\cr
				& NDCG@5 &
				  78.30&78.13&81.35&79.92&78.90 &\textbf{81.86}$^*$\cr
				  
				  & NDCG@10 & 82.32&81.91&84.23&82.71&82.90 &\bf{84.96}$^*$\cr

				\midrule 
				\multirow{3}{*}{ES-EN}& NDCG@1 & 74.88&70.73&74.05&74.55&76.38 &\bf 80.13$^*$\cr
				  & NDCG@5 &
				  75.04&72.34&74.58&75.05&75.10 &\textbf{78.34}$^*$\cr
				  
				  & NDCG@10 & 76.09&74.60&75.99&76.44&76.20 &\bf{78.61}$^*$\cr

			 \cmidrule(lr){2-8}
				\multirow{3}{*}{ES-FR}& NDCG@1 & 67.40&74.97&76.05&77.38&73.97 &\bf 80.21$^*$\cr
				  & NDCG@5 &
				  72.86&74.65&76.75&76.73&75.18 &\textbf{78.97}$^*$\cr
				  
				  & NDCG@10 & 75.51&76.59&78.20&78.16&77.10 &\bf{79.88}$^*$\cr

				 \cmidrule(lr){2-8}
				\multirow{3}{*}{ES-ZH}& NDCG@1 & 64.25& 65.00& 69.35& 65.62&65.75 & \bf 70.70$^*$\cr
				  & NDCG@5 &
				  69.82&68.69&73.16&73.58&70.71 &\textbf{74.75}$^*$\cr
				  
				  & NDCG@10 & 74.08&72.70&75.99&75.85&74.60 &\bf{77.06}$^*$\cr

				\midrule
				\multirow{3}{*}{FR-EN}& NDCG@1 & 71.15&71.28&70.52&76.90&76.23 &\bf 81.03$^*$\cr
				  & NDCG@5 &
				  72.99&72.82&73.99&76.58&75.37 &\textbf{77.73}$^*$\cr
				  
				  & NDCG@10 & 75.46&75.14&75.58&78.03&76.78 &\bf{78.72}$^*$\cr

				  \cmidrule(lr){2-8}
				\multirow{3}{*}{FR-ES}& NDCG@1 & 77.01&74.60&74.43&80.85&78.98 &\bf 83.52$^*$\cr
				  & NDCG@5 &
				  78.18&76.67&77.22&78.89&79.70 &\textbf{80.57}$^*$\cr
				  
				  & NDCG@10 & 79.91&78.41&79.16&80.56&80.81 &\bf{81.69}$^*$\cr

				 \cmidrule(lr){2-8}
				\multirow{3}{*}{FR-ZH}& NDCG@1 & 63.33&62.37&69.75&65.33&65.37 &\bf 70.78$^*$\cr
				  & NDCG@5 &
				  71.73&70.65&73.86&67.82&72.34 &\textbf{74.42}$^*$\cr
				  
				  & NDCG@10 & 75.92&74.49&76.89&74.79&76.16 &\bf{77.47}$^*$\cr
				  
				\midrule
				\multirow{3}{*}{ZH-EN} & NDCG@1 & 56.63&62.83&60.32&61.53&60.45 &\bf 68.52$^*$\cr
                & NDCG@5 &
				  61.69&64.71&64.61&64.53&63.89 &\textbf{68.43}$^*$\cr
				  
				  & NDCG@10 & 64.79&66.99&67.03&66.57&66.43 &\bf{69.72}$^*$\cr
				  
				  \cmidrule(lr){2-8}
				\multirow{3}{*}{ZH-ES}& NDCG@1 & 54.03&59.95&61.55&60.45&63.33 &\bf 67.88$^*$\cr
				 & NDCG@5 &
				  61.64&64.53&66.47&65.61&66.16 &\textbf{68.95}$^*$\cr
				  
				  & NDCG@10 & 66.20&67.99&69.30&68.55&69.19 &\bf{71.09}$^*$\cr

				 \cmidrule(lr){2-8}
				\multirow{3}{*}{ZH-FR}& NDCG@1 & 59.05&53.23&59.97&58.85&59.47 &\bf 65.40$^*$\cr
                & NDCG@5 &
				  63.40&61.68&64.81&63.91&64.84 &\textbf{68.07}$^*$\cr
				  
				  & NDCG@10 & 66.97&65.71&68.34&67.27&68.26 &\bf{70.51}$^*$\cr

				\bottomrule 
			\end{tabular}
			
		\end{threeparttable}
		\caption{NDCG values of baselines and our model. Numbers in the table are in percentages.  * marks statistically significant improvements (t-test with p-value $<$ 0.05) compared with the best baseline.} 
		\label{result} 
	\end{table*}

\subsection{Knowledge Graph}
We use Wikidata~\cite{vrandevcic2014wikidata}, a multilingual KG with entities and relations in a multitude of languages. Up until now, Wikidata contains more than 94 million entities and more than 2000 kinds of relations. And the related entities of queries are annotated by mGENRE~\cite{decao2020multilingual}, a multilingual entity linking model which has a high accuracy of entity linking on 105 languages. 
Table~\ref{know} shows the average number of neighboring entities in each dataset. 

\subsection{Baselines}
To demonstrate the effectiveness of our model, we compare the performance with the following baselines. 

\begin{itemize}
    \item Vanilla BERT~\cite{cedr,clirmatrix}: a fine-tuned multilingual BERT model for CLIR.
    \item CEDR~\cite{cedr}: the contextualized embeddings for document ranking (CEDR) model. This model can be applied to various popular neural ranking models, including KNRM~\cite{knrm}, DRMM~\cite{drmm} and PACRR~\cite{pacrr}, to form CEDR-KNRM/DRMM/PACRR.
    \item HIKE$^{-}$: A variant of HIKE, which concatenates the KG information with the query directly. The difference between HIKE$^{-}$ and HIKE is that HIKE$^{-}$ does not use the hierarchical information fusion mechanism.
\end{itemize}

\begin{table*}[t] 
		\centering 
		\setlength{\abovecaptionskip}{2pt}
		\setlength{\belowcaptionskip}{-5.0pt}
		\fontsize{8}{8}\selectfont    
		\setlength\tabcolsep{3.5pt}
		\begin{threeparttable} 
			\begin{tabular}{lcccccccccccc}  
				\toprule         
				
				\multirow{2}{*}{\bf Model }  & \multicolumn{3}{c}{\bf EN}& \multicolumn{3}{c}{\bf ES}& \multicolumn{3}{c}{\bf FR}& \multicolumn{3}{c}{\bf ZH} \cr

				\cmidrule(lr){2-4}
				\cmidrule(lr){5-7}
				\cmidrule(lr){8-10}
				\cmidrule(lr){11-13}
				& ES & FR & ZH & EN & FR & ZH&EN&ES&ZH&EN&ES&FR \cr
				\midrule 
				HIKE &\textbf{86.18}&\textbf{85.22}&\textbf{85.30}&\textbf{78.61}&\textbf{79.88}&\textbf{77.06}&\textbf{78.72}&\textbf{81.69}&\textbf{77.47}&\textbf{69.72}&\textbf{71.09}&\textbf{70.51}\cr
			    \midrule
				HIKE w/o descriptions & 85.39&84.29&84.05&77.27&79.09&76.35&77.79&80.95&76.41&68.69&70.31&69.51  \cr
				HIKE w/o labels &85.47&84.86&84.81&78.34&79.57&76.38&78.58&81.36&76.71&69.29&70.59&70.34  \cr
				HIKE w/o neighboring entities  &85.33&84.47&84.58&78.03&78.17&76.65&78.15&80.90&76.55&68.65&70.23&69.09\cr
                HIKE w/o target language information &84.68&83.98&83.84&77.70&78.39&76.22&77.79&81.18&76.25&68.59&69.94&69.09 \cr

				\bottomrule 
			\end{tabular}
			
		\end{threeparttable}
		\caption{NDCG@10 of models in ablation study. 
		} 
		\label{abtest} 
	\end{table*}
\begin{figure*}[h]
    \centering
    \setlength{\abovecaptionskip}{1pt}
	  \setlength{\belowcaptionskip}{-10.0pt}

      \includegraphics[width=\linewidth]{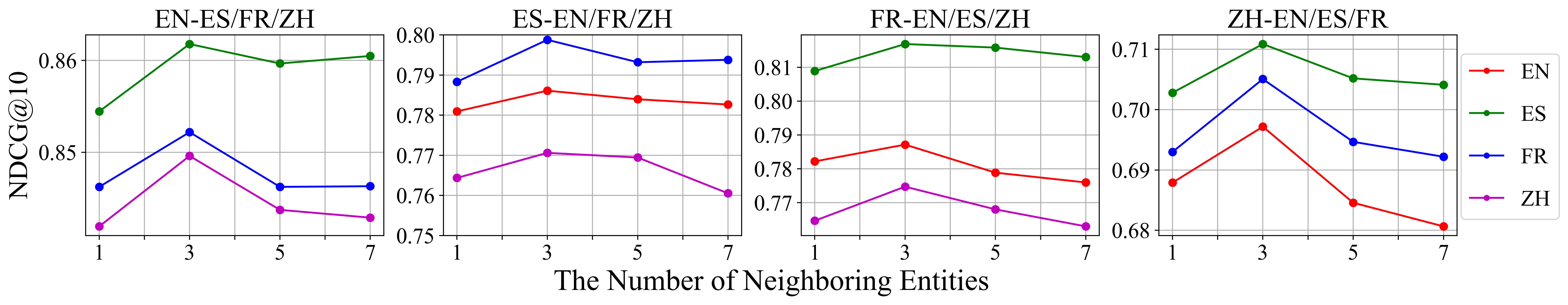}
      \caption{The change of NDCG@10 with the number of neighboring entities increasing.} 
      \label{para}
\end{figure*}

\subsection{Evaluation Metrics}
Normalized Discounted Cumulative Gain~(NDCG) is adopted for evaluation. And we choose NDCG@1, NDCG@5 and NDCG@10~(only evaluate the top 1, 5 and 10 returned documents) as the metrics in all language pairs.  

\subsection{Implementation Details}
In the training stage, the number of heads for the multi-head attention mechanism in knowledge-level fusion is set to 6. In order to reduce the GPU memory and training time, we save the embeddings of entity information before training. The number of all entities we extracted from KG is 376,785. And we only fine-tune the BERT model to obtain textual representations. The learning rates are divided into two parts: the BERT $lr_1$ and the other modules $lr_2$. And we set $lr_1$ to 1e-5 and $lr_2$ to 1e-3. We set the  number of neighboring entities in KG as 3. For those entities without enough neighboring entities, we copy the existing neighboring entities instead. 
We randomly sample 1600 query-document pairs as our training data per epoch. The maximum training epochs are set to 15.

\section{Evaluation Results}

We conduct three experiments to demonstrate the effectiveness of the HIKE model. 

\subsection{Ranking Accuracy}


Table~\ref{result} summarizes the evaluation results of different cross-lingual retrieval models.  From Table~\ref{result}, we have the following findings. 
(i) The results indicate that HIKE significantly and consistently outperforms all the baseline models on 12 language pairs w.r.t all metrics, which demonstrates the effectiveness of the proposed model HIKE. 
(ii) Comparing with Vanilla BERT, the improvement of HIKE$^{-}$ embodies the usefulness and importance of the KG. The external KG makes up for the deficiency of queries and provides  accurate information while ranking the documents. Moreover, the results of HIKE perform better than HIKE$^{-}$, which shows the advantages of our hierarchical fusion mechanism.  
(iii) Specifically, HIKE achieves substantial improvements of both NDCG@1 and NDCG@5 on most datasets comparing with other models, which indicates the knowledge information learned from the entities and neighboring entities is highly related to the task. This result shows that HIKE is capable of ranking the most relevant documents to the top.

All these findings prove that KG information and the hierarchical information fusion can facilitate the CLIR task, and narrow the gap between different languages. 

\subsection{Ablation Study}
In this section, we conduct the ablation study to testify the effectiveness of different information used in HIKE.
In addition, we do the experiments as: 

\begin{itemize}
    \item Remove the labels or descriptions of entities and neighboring entities to verify the effects of them.
    \item Remove the information of neighboring entities to study the influence of neighboring entities.
    \item Remove the information of target language  to learn the importance of them in document ranking.
\end{itemize}

The results are shown in Table~\ref{abtest}. From the results, we observe that 
(i) HIKE obtains the best ranking performance than other incomplete models,  indicating that every part of our model makes contributions to the ranking performance. 
(ii) The model without entity labels outperforms the one without entity description. We conjecture the reason lies in that the information from entity descriptions is more abundant than that from the labels, which is able to provide more beneficial information for the CLIR task.
(iii) The model without target language information performs worst in our ablation test. It demonstrates that target language information plays a significant role in the CLIR task, which establishes an explicit connection between the query in the source language and the documents in the target language.


\subsection{The Effect of Neighboring Entity Number}
In this subsection, we explore the influence of the number of neighboring entities. We set the number of neighboring entities from 1 to 7~(step-size is 2) and conduct the experiments over all datasets. Figure~\ref{para} demonstrates the results, which are divided into four groups according to the different source languages. Each group contains three different target languages.  From the figure, there exists an optimal number of neighbors for each language pair.
The model performance first goes up as the number of neighboring entities increases. After the optimal value, the performance falls down. We conjecture the reason lies in that models with small numbers of neighbors cannot take full advantage of the local neighborhood information in KG, resulting in weak NDCG@10 values. While large numbers of neighboring entities may bring in some unrelated information, leading to unsatisfactory results as well.

\section{Conclusion}
In this paper, we presented HIKE, a hierarchical knowledge-enhanced model for the CLIR task. 
HIKE introduces external multilingual KG into the CLIR task and is equipped with a hierarchical information fusion mechanism to take full advantage of the KG information. 
Specifically, the knowledge-level fusion integrates the KG information in each language. 
And the language-level fusion combines the information from both source and target languages.
The multilingual KG is capable of providing valuable information for the CLIR task, which is beneficial to bridge the gap between queries and documents in different languages.
Finally, extensive experiments on benchmark datasets clearly validated the superiority of HIKE against various state-of-the-art baselines. 


\section{Acknowledgments}
This work is supported by Alibaba Group through Alibaba Innovative Research Program. The research work is supported by the National Key Research and Development Program of China under Grant No. 2017YFB1002104, the National Natural Science Foundation of China under Grant No. U1836206, 62176014, U1811461, and the China Postdoctoral Science Foundation under Grant No. 2021M703273.  Xiang Ao
is also supported by the Project of Youth Innovation Promotion Association CAS and Beijing Nova
Program Z201100006820062.

        

\bibliographystyle{aaai}

\bibliography{Formatting-Instructions-LaTeX-2022.bib}

\end{document}